\newlength{\extralineskip}
\documentclass[12pt]{article}

\begin{document}
\begin{titlepage}
\begin{flushright}
          \begin{minipage}[t]{12em}
          \large UAB--FT--409\\
                 February 1997
          \end{minipage}
\end{flushright}

\vspace{\fill}

\vspace{\fill}

\begin{center}
\baselineskip=2.5em

{\large \bf NEW CONSTRAINTS ON A LIGHT SPINLESS \\
PARTICLE COUPLED TO PHOTONS}
\end{center}

\vspace{\fill}

\begin{center}
{\bf Eduard Mass\'o and Ramon Toldr\`a}\\
\vspace{0.4cm}
     {\em Grup de F\'\i sica Te\`orica and Institut de F\'\i sica
     d'Altes Energies\\
     Universitat Aut\`onoma de Barcelona\\
     08193 Bellaterra, Barcelona, Spain}
\end{center}
\vspace{\fill}

\begin{center}

\large Abstract
\end{center}
\begin{center}
\begin{minipage}[t]{36em}
We obtain new stringent constraints on a light spinless particle $\phi$
coupled only to photons at low energies, considering its effects on
the extragalactic photon background, the blackbody spectrum of
the CMBR and the cosmological abundance of deuterium. 
\end{minipage}
\end{center}

\vspace{\fill}

\end{titlepage}

\clearpage

\addtolength{\baselineskip}{\extralineskip}

\section{Introduction}

Light pseudoscalar (o scalar) particles appear as fundamental 
ingredients of several extensions of the standard model. 
Examples are axions \cite{Weinberg-Wilczek/Peccei} and majorons 
\cite{Chicashige-Gelmini} which arise,
respectively, from the spontaneous breaking of a Peccei-Quinn symmetry   
and of a global lepton symmetry. Other examples are omions \cite{Sikivie}, 
arions \cite{Anselm} and light bosons from extra-dimensional 
gauge theories \cite{Turok}. Pseudoscalar particles normally couple to
photons by means of the Lagrangian:
\begin{equation} \label{lagrange}
  {\cal L} = \frac{1}{8} \ g \phi \ 
    \varepsilon_{\mu \nu \alpha \beta}
                           F^{\mu \nu} F^{\alpha \beta} .
\end{equation}
Axions may be the most famous of them. Their existence would solve
the CP violation problem of QCD and could help to 
explain the composition of the cosmological dark matter. Several
experiments have been or will be performed trying to detect axions.
Many of them are based on the coupling of axions to photons (\ref{lagrange}).
Hence, there is the possibility that these experiments detect
some sort of light bosons different from axions, coupled at low
energies only to photons. 

In Ref. \cite{Masso} the latter possibility was thoroughly explored.
Using laboratory experiments, and several astrophysical and cosmological
observations a set of constraints was found on the parameter space
-coupling constant $g$ and mass $m$- 
of this hypothetical light ($m < 1$ GeV) boson $\phi$ coupled 
at low energy only to photons. Its role as a dark matter component was 
also studied.

Afterwards, further bounds have been published by several 
authors \cite{Mori,Krasnikov,Masso/Brockway,Carugno}.
In particular, using the conversion of
the particles $\phi$ produced in the core of SN 1987A
to photons, on its way to the Earth, by the galactic magnetic field,
one sets the most stringent bound (for $m < 10^{-9}$ eV)
\cite{Masso/Brockway}:   
\begin{equation} \label{prl}
g \leq 3 \times 10^{-12} \ \mbox{GeV}^{-1}.
\end{equation}
This bound relies on data taken by the Gamma-Ray Spectrometer 
on board the satellite SMM, which was on duty when the supernova 
neutrino burst was observed on the Earth. 
Modern spaceborne detectors like EGRET on the Compton GRO 
or the project GLAST
would allow a positive signal, or being pessimistic would set a bound 
more stringent than (\ref{prl}), provided that a type II supernova 
exploded and was observed in our galaxy.

The purpose of this paper is to show that one can also place very 
restrictive constraints on $\phi$ for masses higher than $m\sim 10$ eV,
using three cosmological issues: the extragalactic 
photon background, the cosmic microwave background radiation (CMBR) and the 
cosmological abundances of light elements. 
For certain large regions in the $\phi$ space parameter, the photons
produced by the decay
$\phi \rightarrow \gamma \gamma$ would render a measurable contribution 
to the extragalactic photon background, would induce departures of
the CMBR spectrum from that of a blackbody, and would fission
the deuterium synthesised during the first minutes of the Universe. Using 
experimental data these three facts allow us to set new bounds on $\phi$. 
The present work complements the work presented in \cite{Masso}.

\section{New constraints on $\phi$}

Light bosons $\phi$ are thermally produced
in the early Universe via processes like $e \gamma \rightarrow e \phi$.
Its relic cosmic abundance depends on the freeze-out temperature $T_F$, the
temperature of the primordial plasma when
the rates of production and destruction of $\phi$ became smaller
than the Universe expansion rate. 
Its number density, relative to the number of photons, at time $t$ 
after freeze-out, is given by
\begin{equation} \label{abundance}
\frac{n_\phi}{n_\gamma} = \frac {1}{g_*(T_F)},
\end{equation}
as long as $\tau > t$, being $\tau$ the $\phi$ lifetime. The smaller the
coupling $g$, the higher the temperature $T_F$ and the larger the effective
degrees of freedom $g_*(T_F)$ coupled to $\gamma$ (more annihilations 
to photons have
heated the photons relative to the decoupled $\phi$). We are interested 
in the region $g < 10^{-10}$ GeV$^{-1}$ allowed by astrophysical 
arguments (energy loss in He burning stars and SN 1987A). This small value
of $g$ corresponds to $T_F\gg 200$ GeV. However, we do not know the 
effective degrees of freedom at such high temperature. Clearly, 
some hypothesis must be made to cope with this problem. 
Since our final goal is to constrain the $\phi$ parameters, we adopt
a conservative hypothesis.
Assuming the so-called standard model
desert one obtains $g_*(T_F)\approx 110$, i.e. no more particles appear with
masses larger than 200 GeV (until GUT scales). The prediction of
the supersymmetric standard model desert is $g_*(T_F)\sim 200$. We will
be conservative and assume $g_*(T_F)\sim 500$, which is the
value obtained at $T_F\sim 10^{8}$ GeV 
if one extrapolates by means of an approximate power-law 
the growth rate of the degrees of freedom below 200 GeV. Nevertheless,
the limits we will find are not much sensitive to the precise value of
$g_*(T_F)$. 

The particle $\phi$ is unstable, it decays into two photons with a 
lifetime given by
\begin{equation} \label{tau}
\tau = \frac{64\pi}{g^2 m^3}.
\end{equation}
These decay-produced photons may have different measurable 
consequences for the present state of the Universe depending 
on $z_\tau$, the redshift at which the boson $\phi$ decays
($1+z = R(t=t_0)/R(t=\tau)$ being $t_0$ the age of the Universe
and $R(t)$ the cosmological scale factor). One has to consider four 
important decay epochs:

(1) Lifetimes either $\tau> t_0$ or such that $1 < z_\tau <z_{dec}\approx
1100$, where $z_{dec}$ is
the redshift at which photons last scattered with matter. If $\phi$
decays after matter-radiation decoupling, the decay-produced photons
stream away freely, redshifting due to the expansion of the Universe
(if $\tau<t_0$), and reach the Earth contributing to the photon extragalactic
background. One can compute the present energy flux of photons
per energy and solid angle interval coming from $\phi$ decay (we assume
that $\phi$ is unclustered, which is the least restrictive case):
\begin{equation}
\frac{dF_E}{dEd\Omega} = \frac{n_0}{2\pi \tau H_0}
			\left( \frac{E}{m/2}\right)^{3/2} \exp 
			\left[ \frac{t_0}{\tau} \left(\frac{E}{m/2}
			\right)^{3/2}\right],
\end{equation}
where $H_0$ is the Hubble constant and $n_0$ is the number density
that $\phi$ would have nowadays if it did not decay. 
The measured energy flux can be in general approximated by a power-law
$\propto E^{-\alpha}$ with $\alpha \sim 1$ ($\alpha$ and the normalization
factor depend not very strongly on $E$) \cite{Kolb/Ressell}. The region of
interest falls in the UV part of the spectrum, therefore using
the normalization in the UV region given by \cite{Martin} one obtains
the upper limit
\begin{equation}
\frac{dF_E}{dEd\Omega} < 10^{5} \left( \frac{10\ \mbox{eV}}{E}\right) 
		\ \mbox{cm}^{-2}\mbox{sec}^{-1} \mbox{sr}^{-1}.
\end{equation}
This experimental upper limit constrains the allowed region for the
parameters of $\phi$ as depicted in Fig. 1 (for $\tau > t_0$ see also
\cite{Mori}). We should point out that
this constraint excludes $\phi$ as a component of the cosmological dark 
matter and almost excludes $\phi$ as an important component of the 
galactic dark matter

(2) $z_{rec} < z_\tau <z_{c}\approx 10^5$, being $z_c$ the redshift at
which the rate of Compton scattering $e\gamma \rightarrow e \gamma$ 
becomes too slow to keep kinetic equilibrium between photons and electrons.
The energy dumped by $\phi$ decay heats the electrons
compared to the cosmic background photons. Scattering of these hot electrons
with the cosmic background photons distorts the photon spectrum. The departure
from a blackbody spectrum is parametrized by the Sunyaev-Zeldovich
parameter $y$. Using the Kompaneets equation \cite{Peebles} one
can calculate the relation between $y$ and the energy dumped by $\phi$
decay $\Delta E$:
\begin{equation}
\frac{\Delta E}{E} = e^{4y}-1 \approx 4y,
\end{equation}
where we have advanced that $y$ must be small. Using Eqs. (\ref{abundance})
and (\ref{tau}), energy conservation, and considering an instantaneous
energy release by $\phi$ decay, we obtain
\begin{equation} \label{dumping}
\frac{\Delta E}{E} = 6.6 \left(\frac{500}{g_*(T_F)} \right) 
		\left(\frac{10^{-10} \mbox{GeV}^{-1}}{g} \right) 
		\left( \frac{100\ \mbox{keV}}{m} \right)^{1/2}.
\end{equation}
The COBE FIRAS instrument showed that the CMBR spectrum agrees
with a blackbody spectrum to high accuracy \cite{Mather}. The parameter
$y$ is constrained to be very small, $|y| < 1.5\times 10^{-5}$.
Thus, FIRAS data exclude another region on the plane ($m$, $g$) plotted
in Fig. 1. 

(3) $z_{c} < z_\tau <z_{th} \approx 3\times 10^6$, where at $z_{th}$ the
nonconserving ($e\gamma \rightarrow e \gamma \gamma$) photon number 
processes decouple. At this epoch Compton scattering is fast enough
so that decayed-produced photons can be thermalized. However, since 
photon number cannot be changed one obtains for the CMBR a Bose-Einstein
spectrum, $f=\left(\exp(E/T+\mu)-1 \right)^{-1}$, 
characterized by a nonvanishing photon chemical potential $\mu$,
rather than a blackbody spectrum. Using photon number and 
energy conservation (and assuming that all particles
$\phi$ decay at once at time $\tau$) one gets the following 
expression for the photon chemical potential:
\begin{equation} \label{mu}
\left[ \frac{4}{3}\frac{\zeta (2)}{\zeta(3)}-\frac{\zeta(3)}{\zeta(4)}  
	\right] \mu = \frac{\Delta E}{E} - \frac{8}{3} \frac{1}{g_*(T_F)},
\end{equation} 
for small $\mu$ ($\zeta(n)$ is Riemann's zeta function). 
FIRAS data constrain the photon chemical potential to be $|\mu | < 0.9\times 
10^{-4}$. Making use of Eqs. (\ref{dumping}) and (\ref{mu}), and taking into
account that the contribution to $\mu$ proportional to $1/g_*(T_F)$ is
negligible in the interesting region, we obtain the forbidden zone
plotted in Fig. 1.

(4) At earlier times, the high energy photons coming from $\phi$ decay
create electromagnetic showers that may photofission the deuterium
produced during the first minutes of the Universe. Using the detailed 
analysis of
these showers made in \cite{Sarkar}, we rule out the region depicted
in Fig. 1, which would imply a cosmological deuterium abundance 
D/H $ < 10^{-5}$, smaller than the experimental value. 
Only for masses $m > 10$ MeV, the photon cascades
have enough energy to fission the deuterium nuclei. For masses 
$m > O$(100 MeV), there are also helium photofission and hadronic 
showers that may modify the primordial abundances of light elements, but 
we are not interested in such high masses.  

\section{Conclusions}

A light scalar or pseudoscalar 
$\phi$ coupled at low energies only to photons could contribute
to the extragalactic photon background, could distort the
spectrum of the CMBR and could destroy the primordial deuterium.
The observed extragalactic photon background, the stringent 
limits on the CMBR distortion parameters $y$ and $\mu$ found by
FIRAS, and the measured cosmological abundance of deuterium
allow us to obtain new stringent constraints on $\phi$, extending
the former study carried out in \cite{Masso}.

\newpage

\section*{Acknowledgments}

We thank the Theoretical Astroparticle Network for support under
the EEC Contract No. CHRX-CT93-0120 (Direction
Generale 12 COMA). This work has been partially supported by
the CICYT Research Project Nos. AEN95-0815 and AEN95-0882.
R.T. acknowledges a FPI fellowship from the Ministerio de
Educaci\'{o}n y Ciencia (Spain).

\end{document}